\documentclass[letterpaper]{article}

\usepackage{emulateapj}
\usepackage{onecolfloat}
\usepackage{apjfonts}
\usepackage{amsmath}
\usepackage{natbib}
\usepackage{graphicx}

\newcommand{\Msun}      {\mbox{$\rm\,M_{\mathord\odot}$}}
\newcommand{\Rsun}      {\mbox{$\rm\,R_{\mathord\odot}$}}

\begin{document}

\submitted{Accepted by the Astrophysical Journal}

\twocolumn[
\title{Rotational Broadening Measurement for the Neutron 
Star X-Ray Transient XTE~J2123--058}

\author{John A. Tomsick}
\vspace{0.1cm}
\affil{Center for Astrophysics and Space Sciences, Code
0424, University of California at San Diego, La Jolla, CA,
92093, USA (e-mail: jtomsick@ucsd.edu)}

\vspace{0.2cm}
\author{William A. Heindl}
\affil{Center for Astrophysics and Space Sciences, Code
0424, University of California at San Diego, La Jolla, CA,
92093, USA}

\vspace{0.05cm}
\author{Deepto Chakrabarty}
\affil{Department of Physics and Center for Space Research, 
Massachusetts Institute of Technology, Cambridge, MA 02139, USA}

\vspace{0.05cm}
\author{Philip Kaaret}
\affil{Harvard-Smithsonian Center for Astrophysics, 60 Garden Street, 
Cambridge, MA, 02138, USA}

\begin{abstract}

We used optical spectroscopy of the neutron star X-ray transient 
XTE~J2123--058 to measure the rotational broadening of the binary 
companion's stellar absorption lines and determined that the 
companion's projected rotational velocity is 
$v\sin i = 121^{+21}_{-29}$~km~s$^{-1}$.  This value is considerably 
larger than the measurements of $v\sin i$ obtained previously
for three other neutron star systems where the values are between 34 
and 65~km~s$^{-1}$.  The larger value is likely due to the combination 
of high binary inclination and short (6~hr) orbital period for 
XTE~J2123--058.  Along with previously measured parameters, the 
rotational broadening measurement allowed us to determine the binary 
parameters, including the ratio of the neutron star mass to the 
companion mass, $q = M_{1}/M_{2} = 2.7^{+1.3}_{-0.9}$, the neutron star 
mass, $M_{1} = 1.46^{+0.30}_{-0.39}$\Msun, and the companion mass, 
$M_{2} = 0.53^{+0.28}_{-0.39}$\Msun, assuming a Roche lobe filling 
companion synchronously rotating with the binary orbit.  These values 
are consistent with a previous measurement where the H$\alpha$ emission 
line was used to determine the semiamplitude of the neutron star's 
radial velocity curve ($K_{1}$).  It is a significant result that the 
two methods give consistent values.  We also report the first 
measurement of the XTE~J2123--058 companion's radius.  Assuming 
synchronous rotation, we obtain $R_{2} = 0.62^{+0.11}_{-0.15}$~\Rsun, 
which is in-line with that expected for a late K-type star on or
close to the main sequence.  Finally, we report the first precise 
spectroscopic determination of the binary orbital period ($P_{orb} = 
21442.3\pm 1.8$~seconds).

\end{abstract}

Subject headings: accretion, accretion disks --- stars: neutron --- 
stars: individual (XTE~J2123--058) --- stars: rotation ---
X-rays: bursts --- X-rays: stars

] 

\section{Introduction}

The X-ray transient XTE~J2123--058, discovered in 1998 \citep{lss98},
is a neutron star low-mass X-ray binary (LMXB) with a 6~hr orbital 
period \citep{tomsick99,zurita00}.  Although many of the properties 
of this system are rather typical of LMXBs such as type I X-ray bursts 
and high frequency quasi-periodic oscillations (QPOs), XTE~J2123--058 
distinguishes itself from other LMXBs by having high Galactic latitude 
($b = -36^{\circ}$) and a high and relatively well-determined binary 
inclination ($i = 73^{\circ}\pm 4^{\circ}$; Zurita et al.~2000\nocite{zurita00}).  
A main reason for the good inclination constraint is that partial 
eclipses are present in the outburst optical light curves
\citep{tomsick99,zurita00}.  For optical observations, the high Galactic 
latitude is advantageous since the extinction along the line-of-sight to 
the source is low.  The high binary inclination is useful for measuring 
the rotational broadening because the widths of the companion's absorption
lines increase with inclination.  The ultimate goal of this project
is to obtain a precise measurement of the neutron star mass.

Such mass measurements are important for constraining neutron star 
equations of state (EOS) and for understanding the evolution of neutron 
star systems.  Precise mass measurements have been made for millisecond 
radio pulsars (MSPs, Thorsett \& Chakrabarty 1999\nocite{tc99}), but these 
measurements are lacking for LMXBs.  Since it is theoretically possible to 
spin-up neutron stars in LMXBs to millisecond periods via accretion, 
a link between LMXBs and MSPs has long been suspected \citep{alpar82}.  
Although there is substantial evidence to support this picture, the prediction 
that rapidly rotating neutron stars in LMXBs should be more massive 
by 0.1 to 0.5\Msun~(Bhattacharya 1995\nocite{bhattacharya95}) than those 
that have not been spun-up has not been tested.

Previous work on optical observations of XTE~J2123--058 in quiescence
resulted in measurements of the semiamplitude of the companion's 
radial velocity curve ($K_{2}$), the companion's spectral type, the 
distance to the source and the systemic velocity \citep{tomsick01,casares02}.  
\cite{casares02} also obtained a measurement of the semiamplitude of
the neutron star's radial velocity curve ($K_{1}$) using the 
H$\alpha$ emission line that is present in the spectrum, giving
a determination of the mass ratio ($q = K_{2}/K_{1} = M_{1}/M_{2}$).
Although they considered their $K_{1}$ measurement to be tentative
due to uncertainties concerning using the broad H$\alpha$ line
to infer the radial velocity of the neutron star, they used their 
results to obtain a neutron star mass of $1.55\pm 0.31$~\Msun~(68\% 
confidence errors).  

Here, we use the observations of \cite{tomsick01} to determine the 
projected rotation rate of the companion ($v\sin i$) from a 
measurement of the rotational broadening \citep{mrw94}.  From this, 
we obtain a measurement of $q$ that relies only on measurements of 
the companion's absorption lines and is independent of the 
\cite{casares02} mass ratio measurement.  Although rotational 
broadening measurements have been used to obtain mass measurements 
for several black hole binaries, this has only been previously
accomplished for three other neutron star LMXBs:  Cyg~X-2 \citep{cck98}; 
Cen~X-4 \citep{torres02}; and 2S~0921--630 \citep{shahbaz99}.  A 
measurement of $v\sin i$ was also claimed for Aql~X-1 \citep{scc97};
however, the subsequent discovery of a field star within $0^{\prime\prime}.46$
of Aql~X-1 \citep{wry00} is likely to have some impact on this measurement.
Thus, XTE~J2123--058 provides a relatively rare opportunity to obtain a 
rotational broadening measurement in a neutron star LMXB, which is an 
important step toward obtaining a precise neutron star mass measurement.  
This study also provides a valuable test of the method most commonly 
used to measure compact object masses since we expect to obtain a 
neutron star mass that is considerably less than the 5-15\Msun~compact 
object masses obtained for the black hole systems.

\section{Observations}

In 2000 August, we performed spectroscopy of XTE~J2123--058 with the 
Echelle Spectrograph and Imager (ESI, Sheinis et al.~2000\nocite{sheinis00}) 
on the 10~m Keck II telescope on Mauna Kea, Hawaii.  These observations,
as well as our data reduction techniques, are described in detail by 
\cite{tomsick01}.  We took eight exposures over two nights covering the 
6~hr binary orbit, and an observation log is given in Table~\ref{tab:obs}.  
The dispersion varies from 0.36 to 0.52\AA~per pixel 
(22.8~km~s$^{-1}$~pixel$^{-1}$) over the 4700-6820\AA~band after on-chip 
binning by a factor of two in the dispersion direction.  This is sufficient
to over-sample the spectrograph resolution of 1.0 to 1.5\AA~FWHM, which is 
obtained using a $1^{\prime\prime}$ slit.  Due to the faintness of the 
source ($V$$\sim$$22.5$), relatively long exposure times between 2340 and 
2700 seconds were necessary.  The effective seeing for these exposures varied 
from $0^{\prime\prime}.67$ to $1^{\prime\prime}.23$.  We dereddened the 
XTE~J2123--058 spectra using $A_{V} = 0.37$~magnitudes \citep{hynes01}.  We 
also obtained M, K and G dwarf comparison star spectra with ESI to determine 
the spectral type and to measure the radial velocity curve \citep{tomsick01}.  
In this work, we use K2~V, K4~V and K7~V comparison star spectra.  We used 
the same instrumental setup to obtain these spectra as for XTE~J2123--058 
except that we did not perform on-chip binning for the K7~V comparison star.  
For these short exposures, we obtained seeing between $0^{\prime\prime}.60$
and $0^{\prime\prime}.69$, which is comparable to the seeing for the best
XTE~J2123--058 exposures.  All three comparison star spectra have 
well-measured radial velocities, and we performed a velocity shift 
to produce spectra in the star's rest frame.

\begin{table}[t]
\caption{XTE~J2123--058 Keck Observations\label{tab:obs}}
\begin{minipage}{\linewidth}
\footnotesize
\begin{tabular}{c|c|c|c|c|c} \hline \hline
 & MJD & Exposure & & Orbital & Smear\\
Exposure & (days)\footnote{Modified Julian Date (JD$-$2400000.5) at 
exposure midpoint.} & Time (s) & S/N\footnote{Mean S/N per pixel in
the 5650-5850\AA~band.  Note that these values are slightly lower 
than the values given in \cite{tomsick01} due to a calculation error 
made in that work.  The error does not impact any of the 
\cite{tomsick01} results.} & Phase\footnote{Fraction of the orbit 
since inferior conjunction at exposure midpoint.} & 
(km s$^{-1}$)\footnote{See \S3.1.}\\ \hline
1 & 51759.39442 & 2700 & 1.39 & 0.411 & 195\\
2 & 51759.42822 & 2510 & 2.24 & 0.548 & 205\\
3 & 51759.45771 & 2510 & 1.80 & 0.666 & 108\\
4 & 51759.48768 & 2510 & 3.69 & 0.787 & 52\\
5 & 51759.51748 & 2400 & 3.51 & 0.907 & 172\\
6 & 51759.54536 & 2340 & 2.41 & 0.019 & 199\\
7 & 51759.57288 & 2340 & 2.02 & 0.130 & 137\\
8 & 51760.35770 & 2400 & 5.57 & 0.292 & 55\\
\end{tabular}
\end{minipage}
\end{table}

\section{Results}

\subsection{Rotational Broadening Measurement}

\begin{figure}[t]
\hspace{-1.3cm}
\plotone{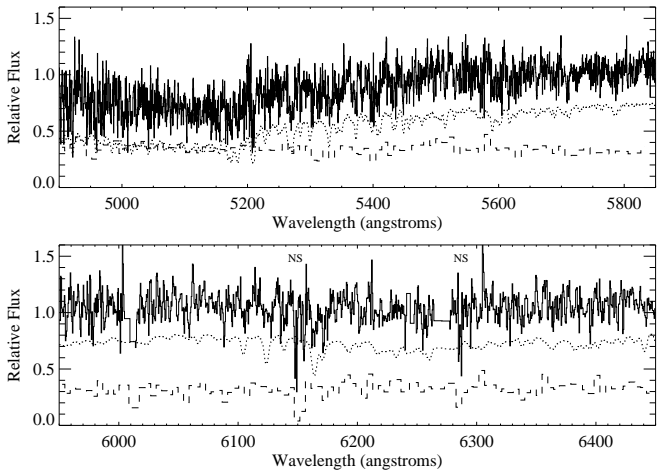}
\vspace{1.0cm}
\caption{The average, Doppler-corrected spectrum of all 
eight XTE~J2123--058 exposures (solid line), covering the
wavelength regions used for the rotational broadening
analysis.  The dotted line is the K7~V comparison star 
spectrum after multiplying by $F = 0.66$ and broadening
by $v\sin i = 121$~km~s$^{-1}$, representing our best 
estimate of the portion of the XTE~J2123--058 spectrum that 
comes from the companion.  The dashed line is the difference
between the spectra for XTE~J2123--058 and the comparison 
star, and these residuals represent the emission from the 
accretion disk.  Features that are likely related to 
imperfectly subtracted night sky (NS) lines are marked.
\label{fig:spectra}}
\end{figure}

The fact that stellar absorption lines are broadened due to rotation 
allows for a measurement of the stellar rotational velocity.  Here, we
used techniques similar to those described in \cite{mrw94}.  We carried 
out the measurement by comparing the average, Doppler-corrected 
spectrum of XTE~J2123--058 in the rest frame of the companion (shown 
in Figure~\ref{fig:spectra}) to the comparison star spectra.  Details 
concerning the production of the XTE~J2123--058 spectrum are provided 
in \cite{tomsick01}.  Previous work \citep{tomsick01,casares02} indicates 
that the spectral type of the XTE~J2123--058 companion is between K5~V 
and K8~V\footnote{\cite{tomsick01} give the range of spectral types as 
K5~V-K9~V, but we have since learned that K9~V is not a valid subclass.}.  
Although we use K2~V, K4~V and K7~V comparison star spectra in this work, 
we expect the K7~V star to provide the most reliable results.  The 
measurement of $v\sin i$ is made by convolving the comparison star 
spectra with rotation profiles from \cite{gray92} for different rotational 
velocities and determining which convolved spectrum provides the best 
match to the XTE~J2123--058 spectrum.  

There are two complications to this technique that must be taken 
into account.  First, the source spectra are smeared since the 
observations each covered a significant fraction of the binary orbit.
To account for this, we determined the smearing and the signal-to-noise
for each of the eight XTE~J2123--058 spectra (see Table~\ref{tab:obs})
and produced eight comparison star spectra convolved with rectangular 
profiles with widths corresponding to the smearing velocities for
the XTE~J2123--058 exposures.  We then calculated the average of the 
eight spectra weighted using the XTE~J2123--058 signal-to-noise 
measurements and convolved this average spectrum with rotation profiles 
for comparison to the average XTE~J2123--058 spectrum.  The effective 
level of smearing prior to the convolution with the rotation profiles
is 100~km~s$^{-1}$.  The second complication is that a fraction
of the light from the source comes from the accretion disk
rather than the companion.  To account for this, we multiplied the 
comparison star spectra by a constant ($F$) prior to minimizing the 
difference between the XTE~J2123--058 spectrum and the comparison 
star spectra.  As both spectra are normalized to 1.0 in the 
5600-5700\AA~band prior to minimization, $F$ corresponds to the 
fraction of the light due to the companion in this band and is
left as a second free parameter in the $\chi^2$-minimization.
It should be noted that this method implicitly assumes that the 
accretion disk contributes a normalized flux of $1$$-$$F$ at all 
wavelengths.

For the rotational broadening measurement, we used the regions
from 4900-5850\AA~and 5950-6450\AA, eliminating regions with 
Balmer series emission lines and also the region around the 
NaI doublet near 5900\AA~where bright night sky lines are 
present.  We broadened the comparison spectra with rotation 
profiles with $v\sin i$ from 5~km~s$^{-1}$ to 200~km~s$^{-1}$ 
in steps of 5~km~s$^{-1}$ and subtracted the broadened 
comparison spectra (multiplied by $F$) from the XTE~J2123--058 
spectrum.  To remove any broadband features that are not relevant 
to the rotational broadening measurement, we divided the 
post-subtraction spectrum into ten wavelength bins, performed 
a cubic spline interpolation of the mean values in these bins 
and subtracted this smooth curve to obtain the residuals.
For each velocity, we calculated residuals for values of $F$
from 0 to 1 in steps of 0.01, found the $\chi^{2}$ deviation 
of the residuals from zero for the 4000 combinations of $v\sin i$ 
and $F$, and used the $\chi^{2}$ values to determine the best
fit parameter values.

\begin{figure}[b]
\plotone{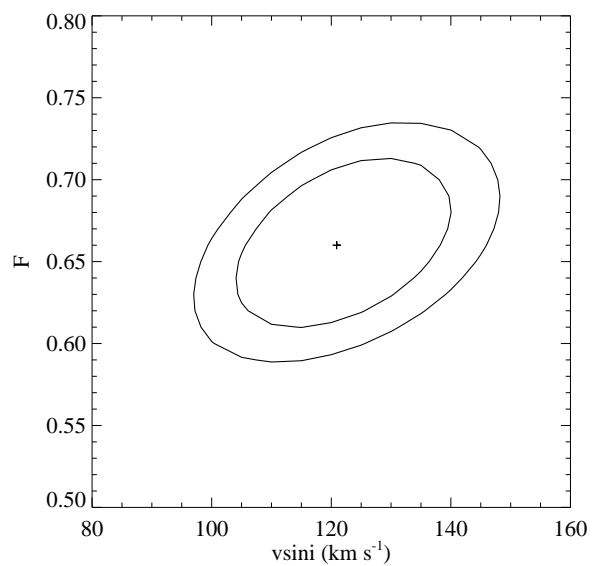}
\caption{Error regions for the two free parameters in the rotational
broadening measurement.  The $v\sin i$ vs. $F$ 68\% and 90\% confidence 
($\Delta\chi^{2} = 2.30$ and 4.61, respectively) contours are shown, 
and the plus marks the best fit values of $v\sin i = 121$~km~s$^{-1}$
and $F = 0.66$.\label{fig:contour}}
\end{figure}

Table~\ref{tab:results} shows the results of the rotational
broadening analysis using the three comparison stars.  In 
addition to the values of $v\sin i$ and $F$ obtained, the 
table includes the linear limb-darkening coefficient 
($\epsilon$) used in broadening the comparison spectra 
(discussed in more detail below) and the values of 
$\chi^{2}/\nu$ for the fit.  As expected, the K7~V comparison 
star provides the best match to the data, and we only include 
the results for the other two comparison stars to point out 
that the value of $v\sin i$ obtained does not depend on which 
comparison star we use.  Figure~\ref{fig:contour} shows 68\% 
and 90\% confidence error contours for $v\sin i$ and $F$ using 
the K7~V comparison star and $\epsilon = 0.6$.  Although the 
contours show that the parameters are slightly correlated, 
the level of correlation is not extreme.  The best fit 
parameter values are $v\sin i = 121^{+19}_{-17}$~km~s$^{-1}$
and $F = 0.66\pm 0.03$ (68\% confidence errors using
$\Delta\chi^{2} = 2.3$).  When performing the $\chi^{2}$-minimization, 
we find reduced-$\chi^{2}$ values somewhat less than 1.0 because 
it is necessary to re-sample the XTE~J2123--058 spectra when the 
Doppler corrections are applied.  The interpolation associated 
with re-sampling means that adjacent wavelength bins are not 
completely independent.  This reduces the rms noise of the 
spectrum, resulting in lower $\chi^{2}$ values.  The most 
conservative approach, which is the one we take, is to permit 
these low $\chi^{2}$ values.  Another approach would be to 
artificially reduce the size of the error bars on the spectrum 
to obtain a reduced-$\chi^{2}$ of 1.0.  However, this would lead 
to a decrease in the parameter error estimates.

\begin{table}[t]
\caption{Rotational Broadening Measurements\label{tab:results}}
\begin{minipage}{\linewidth}
\footnotesize
\begin{tabular}{c|c|c|c|c} \hline \hline
Comparison Star & & $v\sin i$ & & \\
Spectral Type & $\epsilon$\footnote{Linear limb-darkening coefficient.} & 
(km s$^{-1}$) & $F$ & $\chi^{2}/\nu$\\ \hline
K7~V & 0.6 & $121^{+19}_{-17}$ & $0.66\pm 0.03$ & 2620/3333\\
K4~V & 0.6 & $121^{+20}_{-19}$ & $0.76\pm 0.03$ & 2647/3333\\
K2~V & 0.6 & $120^{+21}_{-20}$ & $0.73\pm 0.03$ & 2681/3333\\
K7~V & 0.0 & 115 & 0.66 & 2620/3333\\
K7~V & 0.7 & 123 & 0.66 & 2620/3333\\
\end{tabular}
\end{minipage}
\end{table}

In Figure~\ref{fig:spectra}, the spectrum for the K7~V comparison 
star (dotted line) is shown along with the XTE~J2123--058 spectrum.  
The regions of the spectra shown in the figure correspond to the
regions used in the rotational broadening analysis.  The comparison 
star spectrum has been multiplied by $F = 0.66$ and broadened by 
$v\sin i = 121$~km~s$^{-1}$ so that it represents our best estimate
of the portion of the XTE~J2123--058 spectrum that comes from 
the companion.  The figure also shows the difference between the 
spectra for XTE~J2123--058 and the comparison star, and these 
residuals represent the emission from the accretion disk.  The 
relative flux for the disk spectrum is near 0.3-0.4 at all 
wavelengths, and it should be noted that this occurs because we 
have assumed a flat disk spectrum.  While our method gives a 
value for the companion fraction of $F = 0.66$ in the 
5600-5700\AA~band, Figure~\ref{fig:spectra} indicates that
$F$ varies with wavelength from about 0.5 below 5200\AA~to 
an average value of 0.69 in the 5950-6450\AA~band.  Finally, 
the dips seen in the XTE~J2123--058 spectrum and in the
residuals at 6150\AA~and 6285\AA~are near the wavelengths of
moderately bright night sky lines, and it is likely that 
these dips are due to imperfect background subtraction.
We performed the rotational broadening analysis again after
removing these regions of the spectrum and obtained the
same values for $v\sin i$ and $F$.

We considered possible sources of systematic error in our 
determination of $v\sin i$.  To test our correction for 
smearing, we performed the rotational broadening measurement 
using only exposures 4 and 8, which have the lowest levels
of smearing.  Although the signal-to-noise is lower, the
effective level of smearing is only about 54~km~s$^{-1}$
compared to 100~km~s$^{-1}$ when all eight exposures are
used.  With exposures 4 and 8, we obtained 
$v\sin i = 118^{+21}_{-19}$~km~s$^{-1}$, which is consistent 
with the previously determined value.  This provides evidence 
that the smearing correction does not introduce a systematic 
error significant relative to the statistical errors.  We 
determined that there is no significant systematic error 
associated with dereddening the XTE~J2123--058 spectra by
re-measuring $v\sin i$ with $A_{V} = 0.21$ and 0.53, which 
correspond to the extremes of the error range for the visual 
extinction \citep{hynes01}.  

It is possible that systematic error is introduced due to 
our approximation that the companion is spherical.  \cite{mrw94} 
investigated this for the black hole system A~0620--00 by 
carrying out their analysis with the spherical assumption and 
also by accounting for the shape of the Roche lobe.  They found 
that the spherical assumption introduces a systematic error in 
the measurement of $v\sin i$  below the 2\% level even when 
uncertainties related to gravity darkening are considered.  
This effect is much smaller than our statistical errors and 
should be even smaller for XTE~J2123--058 since the mass ratio 
is less extreme than for A~0620--00.  However, a related concern 
is that $v\sin i$ is dependent on orbital phase for a non-spherical 
companion.  It is likely that this impacts our measurement of 
$v\sin i$ for XTE~J2123--058 because our highest signal-to-noise 
spectra occur near the quadratures (see Table~\ref{tab:obs}).
Based on the orbital variations of $v\sin i$ for the cataclysmic 
variable AE~Aqr \citep{casares96}, we estimate that our measurement 
of $v\sin i$ is not more than 5\% higher than the phase-averaged 
value, and we include this in determining our final error.

Another significant systematic error comes from uncertainties 
in the limb-darkening coefficient ($\epsilon$).  The value of 
$\epsilon = 0.6$ used above is appropriate for a non-irradiated 
K7~V star based on the coefficients from \cite{av99}, and the 
earliest possible spectral type for XTE~J2123--058, K5~V, is 
expected to have $\epsilon = 0.7$.  However, we cannot rule
out the possibility that the correct coefficient values are
lower.  \cite{av99} find that irradiated stars can have 
significantly lower values of $\epsilon$.  In addition, 
$\epsilon$ will not generally be the same for continuum
and line wavelengths, and it is likely that the coefficient 
value will be lower in the lines \citep{ct95}.  Thus, in 
determining the associated systematic error, we consider values 
of $\epsilon$ between 0.0 (no limb-darkening) and 0.7.  As shown 
in Table~\ref{tab:results}, $v\sin i$ varies from 115~km~s$^{-1}$ 
to 123~km~s$^{-1}$ over this $\epsilon$ range.  Adding the 
systematic errors to the statistical error gives our final result 
of $v\sin i = 121^{+21}_{-29}$~km~s$^{-1}$.  

\begin{figure}[t]
\plotone{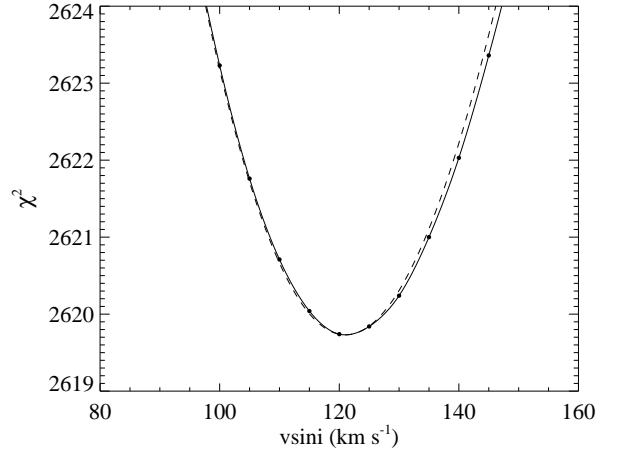}
\caption{Measurement of $v\sin i$ for the XTE~J2123--058
spectrum compared to simulations.  The points show the
difference (in terms of $\chi^{2}$) between the XTE~J2123--058
spectrum and the K7~V comparison star broadened using different
values of $v\sin i$, and the solid line is a cubic spline 
interpolation through the points.  The simulation results 
(dashed line) are in excellent agreement with the measurement
using the actual data.\label{fig:broadening}}
\end{figure}

We performed a final check on our results by carrying out
the rotational broadening measurement using simulated spectra.
We produced simulated spectra by convolving the K7~V 
comparison star spectrum with a 100~km~s$^{-1}$ rectangular 
profile (to simulate smearing) and a 120~km~s$^{-1}$
rotation profile.  We also multiplied the normalized spectrum 
by 0.66 ($F$) and added a constant value of 0.34 ($1$$-$$F$)
to reproduce the situation where not all the light comes from 
the companion star.  Finally, we added noise to the spectrum to 
match the statistics of the XTE~J2123--058 spectrum.  
Figure~\ref{fig:broadening} compares the results of the $v\sin i$ 
measurement for XTE~J2123--058 to the results when the same 
measurement procedure is applied to simulated spectra.  The 
points in the figure show the $\chi^{2}$ values resulting when 
broadened comparison star spectra are used to measure
$v\sin i$ for XTE~J2123--058.  $F$ is left free to minimize 
$\chi^{2}$ for each velocity.  The solid line is a cubic 
spline interpolation through the points.  The dashed line 
shows the result (only the interpolation is shown) when the 
same comparison spectra are used to measure $v\sin i$ for 
the simulated spectra.  The dashed line represents an average 
for 50 simulations.  The agreement between the measurement 
for XTE~J2123--058 and the simulations is excellent, giving 
us confidence that the statistical errors we find for 
XTE~J2123--058 are reasonable.

\subsection{Orbital Period}

Here, we use the epoch of inferior conjunction derived from
our radial velocity curve, HJD~$2451760.0462\pm 0.0013$~days
\citep{tomsick01}, and also from the \cite{casares02} radial
velocity curve, HJD~$2451779.652\pm 0.001$~days, to obtain
the first precise spectroscopic determination of the orbital 
period ($P_{orb}$).  As 79 binary orbits occurred between the 
two measurements, the value for $P_{orb}$ is 
$21442.3\pm 1.8$~seconds (68\% confidence error), which is
approximately 3-$\sigma$ from the value of 
$21447.59\pm 0.15$~seconds derived by \cite{casares02}
from spectroscopic and photometric measurements.  
Although this could be a statistical outlier, \cite{casares02} 
derived $P_{orb}$ by assuming that 2969 binary orbits occurred 
between the photometric epoch of inferior conjunction reported 
in \cite{zurita00} and their spectroscopic epoch.  Based on the 
errors on the epoch and $P_{orb}$ given in \cite{zurita00},
the possibility that 2970 binary orbits occurred can only be 
ruled out at the 1.9-$\sigma$ level, and $P_{orb} = 21440.38\pm 0.15$~seconds 
would be inferred, which is only different from the spectroscopic 
period at the 1.1-$\sigma$ level.  We would conclude that 
this shorter orbital period is correct except for the fact 
that it is 2.2-$\sigma$ from the value of $21445.5\pm 2.3$~seconds 
obtained using photometry only \citep{tomsick99,zurita00}.  
A more accurate measurement of the spectroscopic orbital period
is necessary to determine if either of the $P_{orb}$ values derived 
from spectroscopic and photometric measurements (21440.4 or 21447.6 
seconds) is correct.  A different value could have interesting 
implications for the interpretation of the outburst optical light
curve or possibly the rate of change of the orbital period.
It is desirable to obtain a radial velocity curve by mid-2003 to 
avoid losing the cycle count for the spectroscopic ephemeris.

\section{Binary Parameters and the Neutron Star Mass}

When combined with previous measurements from other work, our 
determination of $v\sin i$ allows us to calculate the parameters of 
the binary system, including the mass ratio, $q = M_{1}/M_{2}$, the 
mass of the neutron star ($M_{1}$) and the mass and radius of the 
companion.  Assuming that the companion fills its Roche lobe and 
rotates synchronously with the binary orbit \citep{paczynski71,mrw94}, 
we use $v\sin i$ along with our previous determination of the 
companion's velocity semiamplitude, $K_{2} = 299\pm 7$~km~s$^{-1}$ 
\citep{tomsick01}, to obtain $q = 2.7^{+1.3}_{-0.9}$.  Using these 
measurements, $P_{orb} = 21442.3\pm 1.8$~seconds and the constraint 
on the binary inclination ($i$) from \cite{zurita00}, we calculate 
the neutron star mass according to
\begin{equation}
M_{1} = \frac{P_{orb}}{2\pi G} \left(\frac{1+q}{q}\right)^2 \frac{K_{2}^{3}}{\sin^{3} i}~~~~~,
\end{equation}
where $G$ is the gravitational constant, leading to 
$M_{1} = 1.46^{+0.30}_{-0.39}$ \Msun~(68\% confidence statistical 
errors and systematic errors).  The companion's mass can also be 
calculated using an expression similar to equation 1, and the 
result is $M_{2} = 0.53^{+0.28}_{-0.39}$~\Msun.  While the mass 
determinations depend on the assumption that the companion fills 
its Roche lobe, $R_{2}$ can be derived from $v\sin i$, $i$ and 
$P_{orb}$ assuming only synchronous rotation, and we obtain 
$R_{2} = 0.62^{+0.11}_{-0.15}$~\Rsun.  This is the first 
measurement of the XTE~J2123--058 companion's radius, and the 
value is in-line with that expected for a late K-type star on
or close to the main sequence.  The radius measurement is also 
relevant to the mass determinations as it provides a check on the 
assumption of a Roche lobe filling companion.  The effective Roche 
lobe radius depends on $M_{2}$ and $P_{orb}$ according to the 
expression given in \cite{casares02}.  Although $M_{2}$ is not 
well-constrained, the effective radius for a 0.5\Msun~companion 
in a binary with a 6~hr orbital period is 0.6\Rsun, which agrees 
with the measured value of $R_{2}$.

Table~\ref{tab:parameters} compares the parameter values
we derive using the rotational broadening measurement to the
values that \cite{casares02} previously obtained by using the 
H$\alpha$ emission line to measure the radial velocity 
semiamplitude of the neutron star ($K_{1}$).  In every case 
where it is possible to compare the two measurements (for 
$q$, $M_{1}$ and $M_{2}$), their best estimate is contained 
within our error range.  To obtain a more direct comparison 
of the two methods, we recalculated the parameters using the 
\cite{casares02} measurement of $K_{2}$ ($287\pm 12$~km~s$^{-1}$).  
These values are given in Table~\ref{tab:parameters} and are 
also consistent with the \cite{casares02} parameters.  We conclude 
that the two measurement techniques give consistent results.  
This is significant for compact object mass measurements in 
general since it provides a test of the (independent) assumptions 
inherent in the two techniques.  However, the comparison between 
the neutron star mass measurements for the two techniques is 
limited to a precision of 20-30\% by the size of the errors.  

For this work and also in \cite{casares02}, the neutron star
and companion mass measurements depend on the determination of 
the binary inclination ($i$) made by \cite{zurita00}.  The 
inclination measurement comes from modeling the outburst optical 
light curves, and it is important to note that the modeling 
depends on $q$ as well as $i$.  For two different light curves, 
\cite{zurita00} find values for the mass ratio between 3 and 5.
While our mass ratio is consistent with the lower part of this
range, the \cite{casares02} value of $q = 2.1\pm 0.4$ lies
somewhat (about 2.3-$\sigma$) below this range.  A mass ratio
as low as the \cite{casares02} value would likely have 
implications for the validity of the inclination measurement.
Thus, it is important to obtain an independent inclination
measurement, and this should be possible through observations
of the ellipsoidal modulations \citep{wd71} in quiescence.

One reason we chose XTE~J2123--058 for this study is that the 
presence of high frequency QPO pairs in the outburst X-ray
emission suggests that the neutron star has been spun up 
to millisecond rotation periods \citep{homan99,tomsick99}.
If neutron stars must accrete significant mass to reach these 
rapid rotation rates, this presents the possibility that an 
massive neutron star might be present that would provide a 
constraint on the neutron star equation of state (EOS).  However, 
our measurement of the neutron star mass is consistent with the 
Chandrasekhar limit of 1.4\Msun, and does not, by itself, constrain 
the EOS.  This work and the work of \cite{casares02} indicate that 
it is unlikely that the neutron star mass in this system is greater 
than 1.8-1.9\Msun.  The precision of the mass measurement from 
this work is limited by statistical errors on the measurement of 
$v\sin i$ (although systematic errors are not negligible), 
indicating that obtaining spectra with better signal-to-noise 
would lead to a tighter mass constraint, which could have 
interesting implications for the amount of mass that must be 
accreted to spin up a neutron star and the long-term accretion 
rate onto the neutron star.  Although XTE~J2123--058 is faint, 
it has the advantage of a rotational broadening signal that is 
considerably larger than for the three other neutron star sources 
where the values of $v\sin i$ are between 34 and 65~km~s$^{-1}$.
The larger value is likely due to the combination of the relatively
high binary inclination and the fact that the XTE~J2123--058 
orbital period is the shortest of the group.  

\begin{table}[t]
\caption{Mass Ratios and Component Masses\label{tab:parameters}}
\begin{minipage}{\linewidth}
\footnotesize
\begin{tabular}{c|c|c|c|c} \hline \hline
Method\footnote{Measurements used to determine the parameters.
Either the rotational broadening (RB) method of this work or 
the method of using the H$\alpha$ emission line to measure 
the radial velocity of the neutron star \citep{casares02}.}
& $K_{2}$ (km~s$^{-1}$)\footnote{The value used for the companion's 
velocity semiamplitude.  Values of $299\pm 7$~km~s$^{-1}$ and 
$287\pm 12$~km~s$^{-1}$ come from \cite{tomsick01} and 
\cite{casares02}, respectively.} & $q$ & $M_{1}$ (\Msun) & 
$M_{2}$ (\Msun)\\ \hline
RB & $299\pm 7$ & $2.7^{+1.3}_{-0.9}$ & 
$1.46^{+0.30}_{-0.39}$ & $0.53^{+0.28}_{-0.39}$\\
RB & $287\pm 12$ & $2.5^{+1.2}_{-0.9}$ & 
$1.35^{+0.32}_{-0.40}$ & $0.54^{+0.32}_{-0.41}$\\
H$\alpha$ & $287\pm 12$ & $2.1\pm 0.4$ &
$1.55\pm 0.31$ & $0.76\pm 0.22$\\
\end{tabular}
\end{minipage}
\end{table}

\section{Summary and Conclusions}

We used moderate resolution optical spectra from Keck Observatory
to carry out a rotational broadening measurement for the 
XTE~J2123--058 companion.  After a detailed analysis where we 
account for possible sources of systematic error, we obtained
$v\sin i = 121^{+21}_{-29}$~km~s$^{-1}$ for the companion's
projected rotational velocity.  Using this result, our
determination of the spectroscopic orbital period and previous 
measurements of $K_{2}$ \citep{tomsick01} and $i$ \citep{zurita00}, 
we calculated the values of the binary parameters, including
$q = 2.7^{+1.3}_{-0.9}$, $M_{1} = 1.46^{+0.30}_{-0.39}$ and
$M_{2} = 0.53^{+0.28}_{-0.39}$, assuming a Roche lobe filling 
companion synchronously rotating with the binary orbit.  Assuming 
only synchronous rotation, we also obtained the first measurement 
of the XTE~J2123--058 companion's radius, 
$R_{2} = 0.62^{+0.11}_{-0.15}$~\Rsun, which is in-line with
that expected for a late K-type star on or close to the
main sequence.

One of the most significant results of this work is that our
measurements of the binary parameters using the rotational
broadening method are consistent with the values found 
by \cite{casares02} by using the H$\alpha$ emission line
to determine $K_{1}$.  Although the comparison of the two 
techniques for obtaining compact object masses is limited 
to a precision of 20-30\% by the errors on $M_{1}$, this
provides an important test of the methods currently 
used to measure compact object masses.  However, there is 
some indication that $q$ is underestimated by the H$\alpha$
emission line method from the fact that \cite{zurita00} obtain 
values of $q$ between 3 and 5 when modeling the XTE~J2123--058
outburst light curves to find the binary inclination.  
While our measurement of $q$ is consistent with this range, 
\cite{casares02} find a lower value of $q = 2.1\pm 0.4$.  
This issue could be explored further by measuring the 
ellipsoidal modulations for XTE~J2123--058 in X-ray 
quiescence.  Also, the error on $v\sin i$ in this work is 
dominated by statistical errors and could be reduced 
through further spectroscopic observations.

\acknowledgements

We wish to thank Jeffrey McClintock, John O'Meara and Dawn 
Gelino for useful suggestions and discussions.  We acknowledge 
Bob Goodrich for his help with ESI.  We thank Gary Puniwai for 
assistance.  We acknowledge the referee for finding a calculation
error in the first version of this paper and for other useful
comments.  Data for this work were obtained at the W.M. Keck 
Observatory, which is operated as a scientific partnership among 
the University of California, Caltech and NASA.


\end{document}